# Identificación de nuevos medicamentos a través de métodos computacionales.

Raúl Isea*

Fundación Instituto de Estudios Avanzados IDEA, Centro de Biociencias, Valle de Sartenejas, Baruta 1080, Venezuela. Email: risea@idea.gob.ve

**Resumen**.

El desarrollo de nuevos medicamentos es un problema complejo que carece de una solución única y automática desde un punto de vista computacional, debido a la carencia de programas que permitan manejar grandes volúmenes de información que están distribuidos a lo largo de todo el mundo entre múltiples bases de datos. Por ello se describe una metodología que permita realizar experimentos *in silico* para la identificación actual de nuevos medicamentos.

**Palabras Claves**: Medicamento; Experimentos *in silico*; Docking; Tecnología de la Información.

## 1. Introducción.

Gracias a los avances en la automatización de los procesos biotecnológicos se ha observado en los últimos veinte años un crecimiento exponencial en el número de secuencias genómicas que están disponibles tanto por la comunidad académica como científica (figura 1). Si se consideramos el volumen de información que está almacenada en una sola base de datos como por ejemplo, el Centro Nacional para la Información Biotecnológica (conocido por su siglas en inglés NCBI que significa "*National Center for Biotechnology Information*"), donde están almacenadas 101.815.678 secuencias genómicas pertenecientes a diversos organismos y provenientes de varios laboratorios internacionales (1). Dicho conjunto de secuencias están compuestos por más de ciento un mil millones de nucleótidos hasta febrero de 2009 (para ser exacto: 101.467.270.308). Para tener una dimensión clara del volumen de la información que ello representa, considérese por ejemplo el trabajo científico titulado: "*Identification and analysis of functional elements in 1% of the human genome by the ENCODE pilot project*", liderizado por el Instituto Nacional del Genoma Humano (2). Dicho trabajo está escrito con 119.465 caracteres (sin espacio) en 18 páginas diferentes. Sí se considera una "posible equivalencia" donde cada carácter de dicho trabajo represente un nucleótido (a modo ilustrativo), entonces el volumen de información que se consulta en el NCBI puede ser "equivalente" a un libro compuesto por más de dieciséis millones de folios diferentes.

Más aún, en los últimos años debido al crecimiento de la información y el auge de los repositorios que están ubicados a nivel mundial, se está fomentando bases de datos de base de datos que completan una gama de información que está disponible por la comunidad científica, como por ejemplo: *Database of Biological Database* disponible a través de la página Web ubicada en http://www.biodbs.info/

Por los comentarios anteriores, se debe solventar la brecha tecnológica que existe actualmente para obtener la información en el menor tiempo posible. Sí además se desea realizar cualquier operación que involucre, por ejemplo, buscar cuán similares son las secuencias que se han depositado en bases de datos (conocido como alinear secuencias), los tiempos de cómputo se incrementan significativamente. Para visualizar esta última afirmación, considérese el cálculo que se realizo en el Laboratorio Nacional de Argonne donde se dispuso una supercomputadora donde se dedicaron más de diez mil procesadores en un cálculo simultáneamente (3), con el objetivo de identificar todas aquellas secuencias de los genomas microbianos que sean similares entre sí, y puedan identificar aquellas secuencias que aún no se hayan anotado hasta el momento. Dicha comparación generó Petabyte de datos, es decir, cercano al millón de GB de datos o el equivalente de espacio que ocuparían veinticinco mil discos duros de 40 GB cada uno. En este sentido es necesario innovar en el desarrollo de procesos computacionales que estén dirigidos a procesar cualquier tipo de información en forma eficiente e independientemente del volumen de datos que se requiera analizar (en la figura 2 se muestra un ejemplo de una pequeña sección que corresponde al gen *rpoB* del resultado de este alineamiento).

Por ello, el presente trabajo tiene como objetivo presentar una metodología computacional para la identificación de nuevos medicamentos señalando los requerimientos computacionales necesarios para ello. Asimismo se resalta el gran poder de cómputo científico requerido en algunos ejemplos encontrados en la literatura científica solo disponible en centros de computación avanzados.

Finalmente acotar que este trabajo destaca algunas estrategias para la identificación de potenciales medicamentos por métodos computacionales, pero no intenta descartar otras aproximaciones como por ejemplo QSAR (4).

## 2. Actualización en la Bibliografía y selección del problema.

El primer paso para el diseño de nuevos fármacos por métodos computacionales es la identificación de aquellas proteínas que puedan ser consideradas como blanco de interacción por una determinada droga. Para ello se requiere hacer una búsqueda bibliográfica de todos los trabajos relacionado en dicha enfermedad. En este sentido se suele emplear la base de datos del NCBI (disponible en la página web www.ncbi.nlm.nih.gov) que posee un registro de las diferentes publicaciones científicas en todo el mundo (centradas en el área de salud). Sí se realiza una búsqueda en dicha base de datos cuyo título en el trabajo posea la palabra droga, obtenemos un poco menos de cuatro millones de trabajos publicados hasta noviembre de 2007 (exactamente 3.777.179) en 134 revistas diferentes. Este número de referencias nos da una idea general del volumen de información que está disponible en el tema, y comprender porque algunos laboratorios en todo el mundo se especializan en una sola afección.

De hecho, sí se realizan nuevas búsquedas de trabajos científicos empleando como palabras 'droga con alguna enfermedad' como por ejemplo "droga y tuberculosis", "droga y Chagas", "droga y dengue", "droga y cáncer", "droga y HIV"; el número de trabajos publicados hasta noviembre de 2007 solo asciende a 59.518, 2.805, 1.490, 533.293, y 89.924 respectivamente. Como se puede apreciarse a partir de dichos valores, se ha obtenido más resultados en solventar el problema de cáncer y sida, y posteriormente en combatir la tuberculosis (demás está decir que este último comentario es cierto, sí y solo sí nuestro universo fuera únicamente estas enfermedades).

Sí se considera el caso de la malaria, por centrar este trabajo en un ejemplo, el número de publicaciones hasta noviembre de 2008 asciende solo a 22.761, es decir, menos del uno por cierto de todas las publicaciones que se han realizado en esta área de trabajo. Recordemos que la malaria o paludismo es una enfermedad tropical que hasta la fecha se conocen cuatro cepas diferentes que afecta a la especie humana, donde *Plasmodium falciparum* es la cepa con el mayor índice de incidencia, llegándose a registrar 500 a 800 millones de casos anuales aproximadamente en todo el mundo. En el mapa mundial que está disponible en la figura 3, donde se muestra las zonas que están afectadas por Malaria y aquellas zonas que están en riesgo de ser afectadas por la cepa de *Plasmodium falciparum* de acuerdo con datos obtenidos de la Organización Mundial de la Salud hasta el 2007; mientras que en Latinoamérica, la cepa que representa el mayor número de casos es la de *Plasmodium vivax*, donde se han obtenido más de 80 millones de casos de acuerdo a datos suministrados por la Organización Panamericana de Salud (OPS).

Por otra parte se debe tener presente que la Organización Mundial de la Salud está auspiciando una campaña a nivel mundial para erradicar dicho flagelo a partir de 1998, titulado: "*Hacer retroceder la malaria*" para el 2015 (5). Para lograr ello, se están invirtiendo 3.000 millones de dólares. Quizás se pueda considerar que esta inversión es realmente significativa, pero recordemos que se han gastado aproximadamente 30 mil millones de dólares entre gastos médicos y por pérdidas económicas por concepto de dicha enfermedad únicamente en África. No obstante, como recientemente se indicó en la Organización de las Naciones Unidas, se requiere para el 2010 otros 3.000 millones de dólares para continuar el plan de erradicación propuesto por ellos (detalles en la página Web http://www.unicef.org/spanish/health/index_malaria.html).

## 3. Identificación de las potenciales proteínas blanco a drogas.

Una vez revisada la literatura, el siguiente paso es identificar todas aquellas proteínas que han sido citadas en la misma, el cual constituye un paso crítico en la investigación. Sin embargo, a pesar que el número de trabajos que se han publicado en el área de búsqueda de potenciales candidatos a blanco en malaria es pequeño, se debe centrar dicho estudio en una cepa en particular. En este caso se considero la de *Plasmodium falciparum* al ser la más virulenta. Por lo que se ha identificado más de una centena de posibles proteínas candidatas a blanco que puede interactuar con una droga específica. Para visualizar el número posible de proteínas a estudiar deberíamos incluir a todas aquellas asociadas a 1) procesos de inhibición de proteasas, 2) involucradas en la vía de síntesis de ácidos grasos, 3) las que interfieren la biosíntesis d'isoprenoides, entre otras más. Por ello se está desarrollando algoritmos computacionales que permitan ensayar búsquedas automáticas con ayuda de un lenguaje de programación de alto nivel llamado Python, el cual está disponible gratuitamente en la página Web www.python.org (actualmente se está preparando un trabajo científico de ello).

## 4. La necesidad de integrar datos con ayuda de Workflows.

No basta con emplear una base de datos de proteínas como la del NCBI sino además se deben incluir otras bases de datos donde se hayan considerado otras secuencias potenciales a blanco a droga, como por ejemplo las bases de datos Swissprot (disponible en la página Web www.expasy.org/sprot/), PIR (http://pir.georgetown.edu/pirwww/dbinfo/pir_psd.shtml), Unipro (www.uniprot.org/), entre otras más. Por ello ésta integración, y su posterior verificación de la misma, genera otro problema tecnológico como es la unificación de todos estos datos en una misma plataforma computacional. En ese sentido, se está desarrollando Workflow (palabra inglesa que significa "*Flujo de Trabajo*"), es decir, programas computacionales especialmente diseñados para manejar la información en forma automática como es el caso del programa Taverna (destalles en http://taverna.sourceforge.net/). La ventaja que posee este programa que está desarrollado bajo una filosofía de trabajo de Software Libre y a su vez, permite impulsar la cooperación internacional para encontrar soluciones a problemas de interés común. En este sentido, se está comenzando a implementar PhyloGrid por la comunidad académica que permiten realizar cálculos automáticos en filogenia molecular con ayuda del programa MrBayes (6-10).

## 5. Docking computacional, un posible paso para la búsqueda de probables de medicamentos.

El último paso que permitirá hacer una elección de potenciales fármacos que probablemente puedan inhibir o activar una respuesta biológica capaz de ser empleado para combatir una enfermedad específica, es la técnica computacional llamada Docking. Esta metodología consiste en cuantificar la interacción entre la droga y la proteína empleando cálculos de energía de interacción. Para lograr ello, se debe ensayar una amplia gama de posibles posiciones aleatorias de la droga alrededor de la proteína, y en cada caso se debe calcular la energía de interacción resultante de la misma, y posteriormente se selecciona aquella conformación que presenta el menor valor de energía producto de la misma. Esta demás señalar que dichos cálculos estarán sesgados por la elección de la droga o familias de drogas que se empleen en dicho estudio. Para evitar ello, se deben emplear bases de datos de drogas como por ejemplo ZINC, disponible en la página Web http://zinc.docking.org para que de esa manera se favorezca el descubrimiento de nuevos medicamentos útiles contra una determinada enfermedad (actualmente dicha base de datos cuenta con un poco más de seis millones de drogas en sus registros).

Por otra parte existe una gama de programas computacionales que permiten realizar cálculos del tipo Docking entre los que podemos mencionar: AutoDock (disponible en http://autodock.scripps.edu/), ZDOCK – RDOCK (disponible en http://zlab.bu.edu/zdock/index.shtml), DOCK (http://dock.compbio.ucsf.edu/), DOT (http://www.sdsc.edu/CCMS/DOT/), FTDock - RPScore y MultiDock (http://www.sbg.bio.ic.ac.uk/docking/), GRAMM (http://vakser.bioinformatics.ku.edu/resources/gramm/gramm1/), Hex (http://www.csd.abdn.ac.uk/hex/), y mucho más. La selección de dichos programas es por sí un trabajo de investigación especializado, pero a continuación se muestran tres ejemplos publicados en la literatura científica donde se empleo algunos de ellos destacando los recursos computacionales empleados en los mismos.

En primer lugar, el cálculo que se realizara en el supercomputador Blue Gene de la empresa IBM que hasta la fecha es el supercomputador más rápido del mundo, de acuerdo a los resultados obtenidos en la página Web top500.org. Dicho cálculo se empleo el programa DOCK, donde se dedicaron 16.384 procesadores del tipo PowerPC (440 dual-core y 1 GB de RAM por procesador). Ellos realizaron el cálculo para identificar aquellas posibles drogas que pueden ser efectivas contra el HIV-1 (11).

En segundo lugar está el cálculo que se realizara en el supercomputador de San Diego (conocido por su sigla en inglés SDSC) para la identificación de nuevos medicamentos contra la gripe aviar (12). En dicho proceso se identificaron 27 potenciales fármacos similares a Tamiflu con ayuda del programa AutoDOCK en 128 procesadores del tipo PowerPC.

El último ejemplo es el cálculo que realizara la iniciativa internacional denominada WISDOM (abreviatura de las siglas en inglés que significa: "*Wide In-Silico Docking Of Malaria*") para la búsqueda de nuevas drogas contra la malaria (13), donde se empleo el programa AutoDOCK. A pesar que dicho cálculo se realizo en varios cientos de procesadores, dicho cálculo tardó 76 días consecutivos en una plataforma de cálculo distribuido entre diversos países del mundo (14-17). De manera que salvando todos estos detalles comentados a lo largo del trabajo, puede decirse que es posible dedicarse a la búsqueda de nuevos medicamentos donde el computador es justamente la herramienta experimental que hace posible ello.

## 6. Limitación en estas metodologías computacionales.

A pesar del esfuerzo computacional invertido por diversos centros de computación, la principal limitación de emplear estas metodologías computacionales es la gran cantidad de posibles soluciones posibles que se generan las mismas, conocidas como falsos-positivos; debido a que el resultado de la evaluación de la interacción energética droga-proteína al estar relativamente cercana entre sí (información adicional en la página http://shoichetlab.compbio.ucsf.edu/docking.php). En el caso de la iniciativa internacional WISDOM (14), el volumen de la información que se genero fue del orden de 1600 Gigabytes de datos, imposible de analizar manualmente (ver figura 4). En este caso se está manejando una herramienta computacional de distribución gratuita conocida como DIANE (detalles en la página Web http://cern.ch/diane) especialmente desarrollada para trabajar en entornos computacionales del tipo computación Grid.

Por ello y hasta la fecha, se está invirtiendo un gran esfuerzo en encontrar una función energética que disminuya el número de falsos positivos, y probablemente sea considerando parámetros físico-químicos que estén estrechamente ligados a la respuesta de la red metabólica de la proteína blanco involucrada. Esta última afirmación proviene del hecho que la proteína seleccionada no está aislada sino que obedece a una respuesta celular determinada.

Para finalizar, es importante reseñar el trabajo publicado por Mahdavi y Lin (18) donde dichos autores emplearon una técnica de ontología para reducir el número de posibles resultados que se generan con la metodología Docking descrito *grosso modo* en este trabajo.

## 7. Conclusiones.

El presente trabajo destaca la necesidad de innovar en el desarrollo de plataformas computacionales que integren la información que proviene de diversos centros e instituto de investigación dedicados a la búsqueda de nuevos medicamentos que ayuden a mejorar la calidad de vida. Asimismo se destaca la necesidad de desarrollar nuevas metodologías computacionales que permitan encontrar nuevas funciones energéticas que permitan reducir el número de falsos positivos que se generan con el método computacional Docking. Por todo ello, es importante resaltar la cooperación interdisciplinaria para nuclear soluciones noveles que pueden ser empleadas en este tipo de problemas (como la iniciativa internacional WISDOM). Demás esta señalar que los datos generados por estas metodologías computacionales permitirán reducir el número de estudios experimentales *in vitro* necesario para validar los resultados encontrados por métodos teóricos.

## 8. Agradecimientos.

## Dedicatoria.

Este trabajo está dedicado a la memoria de Leonardo Mateu Suay.

## Referencias.

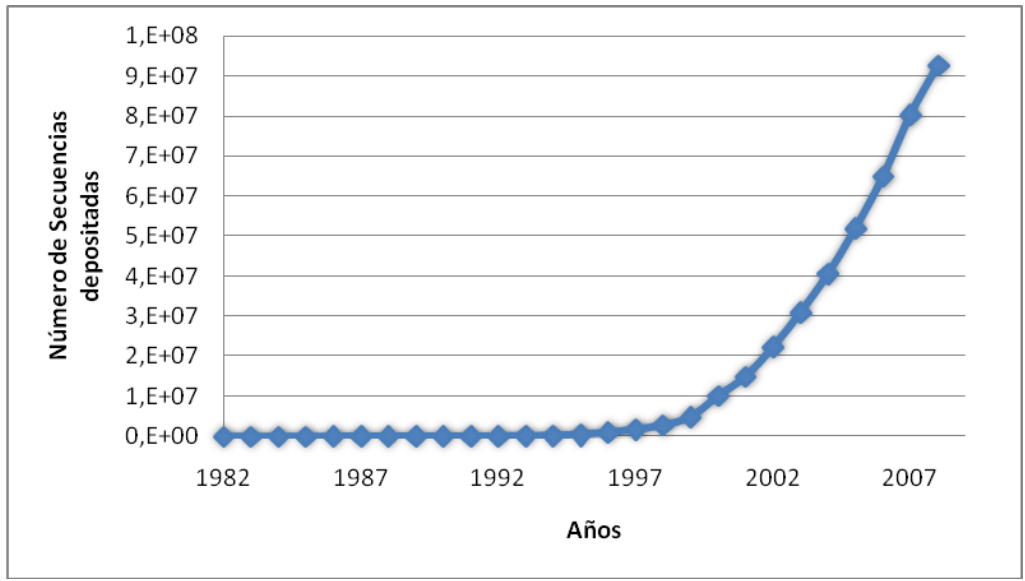

**Figura 1**. Estadística anual del número de secuencias almacenadas en la base de datos del NCBI, disponible a través del sitio web http:// www.ncbi.nlm.nih.gov

```
NC_000962    TTGGCAGATTCCCGCCAGAGCAAAACAGCCGCTAGTCCTAGT
CP000611     TTGGCAGATTCCCGCCAGAGCAAAACAGCCGCTAGTCCTAGT
AE000516     TTGGCAGATTCCCGCCAGAGCAAAACAGCCGCTAGTCCTAGT
BX842574     TTGGCAGATTCCCGCCAGAGCAAAACAGCCGCTAGTCCTAGT
AP010918     TTGGCAGATTCCCGCCAGAGCAAAACAGCCGCTAGTCCTAGT
CP000717     TTGGCAGATTCCCGCCAGAGCAAAACAGCCGCTAGTCCTAGT
BX248336     TTGGCAGATTCCCGCCAGAGCAAAACAGCCGCTAGTCCTAGT
L27989       TTGGCAGATTCCCGCCAGAGCAAAACAGCCGCTAGTCCTAGT
U12205       TTGGCAGATTCCCGCCAGAGCAAAACAGCCGCTAGTCCTAGT
CP000854     -------------------------ACAGACGCTAGTCTCCAT
CP000325     -------------------------ACAGACGCTAGTCTCCAT
FM211192     TTGCCAGATTTCGGCCAGAGCAAGACAGACGTTAGTCCTAGC
AL583923     TTGCCAGATTTCGGCCAGAGCAAGACAGACGTTAGTCCTAGC
Z14314       TTGCCAGATTTCGGCCAGAGCAAGACAGACGTTAGTCCTAGC
EU203594     TTGCCAGATTTCGCCCAGAGCAAGACAGGCGTTAGTCCTAGC

NC_000962    CCGAGTCGCCCGCA
CP000611     CCGAGTCGCCCGCA
AE000516     CCGAGTCGCCCGCA
BX842574     CCGAGTCGCCCGCA
AP010918     CCGAGTCGCCCGCA
CP000717     CCGAGTCGCCCGCA
BX248336     CCGAGTCGCCCGCA
L27989       CCGAGTCGCCCGCA
U12205       CCGAGTCGCCCGCA
CP000854     CAGGGTCGCCCGCA
CP000325     CAGGGTCGCCCGCA
FM211192     CAGAGTCGCCCCCA
AL583923     CAGAGTCGCCCCCA
Z14314       CAGAGTCGCCCCCA
EU203594     CAGAGTCGTCCCCA
```

**Figura 2**. Se muestra el alineamiento obtenido de una pequeña sección del alineamiento obtenido de los genomas del tipo microbiano que pertenece únicamente al gen *rpoB*. Finalmente, señalar que un alineamiento de tan solo 100 genomas del tipo mycobacterium (por citar un ejemplo), solo el volumen de información inicial es de 178 MB en disco duro, y generará varias gigabytes de datos que deben analizarse posteriormente (a modo de comentario personal).

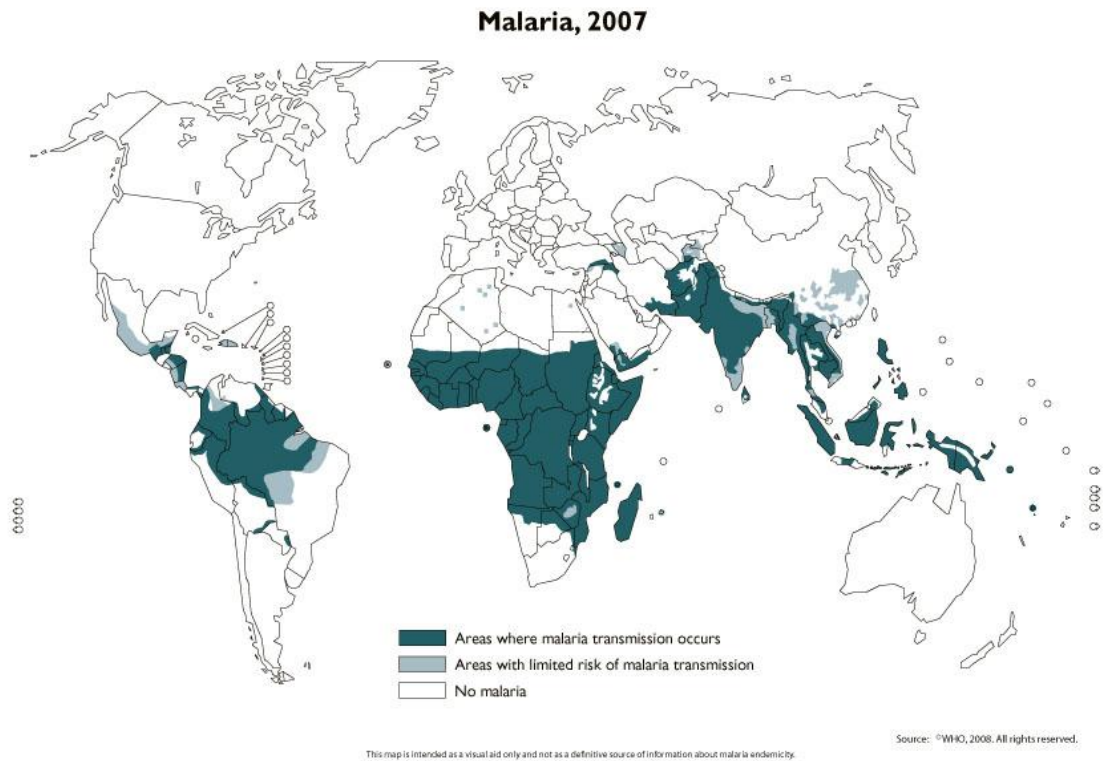

**Figura 3**. Mapa Mundial donde se muestra las zonas que están afectadas por Malaria y aquellas zonas que están en riesgo de ser afectadas por la cepa de *Plasmodium falciparum*. Dicha figura se descargo de la página Web http://www.who.int/ith/maps/en/

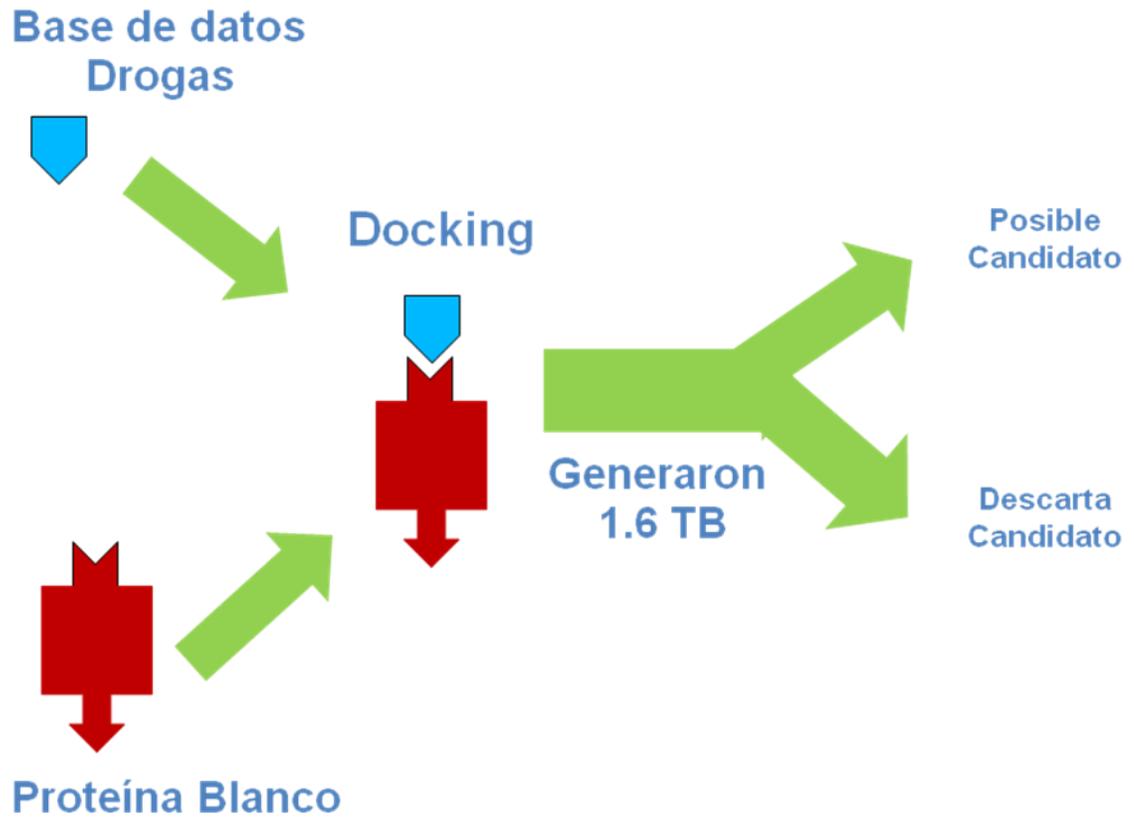

**Figura 4**. Esquema general del proceso computacional para la búsqueda de nuevos medicamentos de acuerdo al protocolo descrito en el proyecto internacional WISDOM-II (17), donde se observa que una proteína Blanco (de color rojo) es la proteína blanco de diversas drogas que están en la Base de Datos de medicamentos llamado ZINC (ver la sección 5). El próximo paso es el cálculo del Docking propiamente dicho, donde se llego a generar 1600 GB de datos cuando se empleo un blanco útil contra la malaria. Finalmente indicar que el procesamiento de los resultados se basa en seleccionar o no aquellos compuestos que pueden ser candidatos a medicamento contra la misma.